\documentclass[usenatbib]{mn2e}
\usepackage{psfig}
\usepackage{times}
\usepackage{amssymb}
\usepackage{graphicx}
\usepackage{graphics}

\def\gs{\mathrel{\raise0.35ex\hbox{$\scriptstyle >$}\kern-0.6em\lower0.40ex\hbox{{$\scriptstyle \sim$}}}}
\def\ls{\mathrel{\raise0.35ex\hbox{$\scriptstyle <$}\kern-0.6em\lower0.40ex\hbox{{$\scriptstyle \sim$}}}}
\def\kms{\,\hbox{km}\,\hbox{s}^{-1}}
\def\Msol{\mathrel{\rm M_{\odot}}}

\def\Msolyr{\mathrel{\rm M_{\odot}\,yr^{-1}}}
\def\Wm2{\,\hbox{W}\,\hbox{m}^{-2}}
\def\gsim{\mathrel{\raise0.35ex\hbox{$\scriptstyle >$}\kern-0.6em\lower0.40ex\hbox{{$\scriptstyle \sim$}}}}
\def\lsim{\mathrel{\raise0.35ex\hbox{$\scriptstyle <$}\kern-0.6em\lower0.40ex\hbox{{$\scriptstyle \sim$}}}}
\def\ltsima{$\; \buildrel < \over \sim \;$}
\def\simlt{\lower.5ex\hbox{\ltsima}}
\def\gtsima{$\; \buildrel > \over \sim \;$}
\def\simgt{\lower.5ex\hbox{\gtsima}}
\def\cgs{\mathrel{\rm erg\,cm^{-2}\,s^{-1}}}

\begin{document}

\title[Swinbank et al.]{A Spatially Resolved Map of the Kinematics,
  Star-Formation and Stellar Mass Assembly in a Star-Forming Galaxy at
  $z$=4.9}

\author[]
{ \parbox[h]{\textwidth}{
A.\ M.\ Swinbank\,$^{\! 1\, *}$,
T.\ M.\ Webb\,$^{\! 2}$,
J.\ Richard\,$^{\! 1}$
R.\,G.\ Bower\,$^{\! 1}$,
R.\ S.\ Ellis\,$^{\! 3}$,
G.\ Illingworth\,$^{\! 4}$,
T.\ Jones\,$^{\! 3}$,
M.\ Kriek\,$^{\! 4}$,
I.\ Smail\,$^{\! 1}$,
D.\ P.\ Stark\,$^{\! 5}$
\& P.\ van Dokkum\,$^{\! 6}$
}
\vspace*{6pt} \\
$^1$Institute for Computational Cosmology, Durham University, South Road, Durham, DH1 3LE, UK \\
$^2$McGill University, Department of Physics, 3600 rue University, Montreal, QC H2A 2T8, Canada \\
$^3$Astronomy Department, California Institute of Technology, 105-24, Pasadena, CA 91125, USA \\
$^4$UCO/Lick Observatory, University of California, Santa Cruz, CA 95064, USA \\
$^5$Institute of Astronomy, University of Cambridge, Madingley Road, Cambridge, CB3 0HA, United Kingdom\\
$^6$Department of Astronomy, Yale University, New Haven, CT 06520, USA \\
$^*$Email: a.m.swinbank@durham.ac.uk \\
}

\maketitle
\begin{abstract}
  We present a detailed study of the spatially resolved kinematics,
  star-formation and stellar mass in a highly amplified galaxy at
  $z=4.92$ behind the lensing cluster MS\,1358+62.  We use the observed
  optical, near- and mid-infrared imaging from {\it HST} ACS \& NICMOS
  and {\it Spitzer} IRAC to derive the stellar mass and the Gemini/NIFS
  IFU to investigate the velocity structure of the galaxy from the
  nebular [O{\sc ii}]$\lambda\lambda$3726.8,3728.9 emission.  Using a
  detailed gravitational lens model, we account for lensing
  amplification factor 12.5$\pm$2.0 and find that this intrinsically
  L$^{*}$ galaxy has a stellar mass of
  M$_{\star}$=7$\pm$2$\times$10$^{8}$M$_{\odot}$, a dynamical mass of
  M$_{dyn}$=3$\pm$1$\times$10$^{9}${\it csc$^2$(i)}\,M$_{\odot}$
  (within of 2\,kpc) and a star-formation rate of
  42$\pm$8\,M$_{\odot}$\,yr$^{-1}$.  The source-plane UV/optical
  morphology of this galaxy is dominated by five discrete star-forming
  regions.  Exploiting the dynamical information we derive masses for
  individual star-forming regions of M$_{cl}\sim$10$^{8-9}$M$_{\odot}$
  with sizes of $\sim$200\,pc.  We find that, at a fixed size, the
  star-formation rate density within these H{\sc ii} regions is
  approximately two orders of magnitude greater than those observed in
  local spiral/starburst galaxies, but consistent with the most massive
  H{\sc ii} regions in the local Universe such as 30\,Doradus.
  Finally, we compare the spatially resolved nebular emission line
  velocity with the Ly$\alpha$ and UV-ISM lines and find that this
  galaxy is surrounded by a galactic scale outflow in which the
  Ly$\alpha$ appears redshifted by $\sim$150\,km\,s$^{-1}$ and the
  UV-ISM lines blue-shifted by $\sim$-200\,km\,s$^{-1}$ from the
  (systemic) nebular emission.  The velocity structure of the outflow
  mirrors that of the nebular emission suggesting the outflow is young
  ($\lsim$15\,Myr), and has yet to burst out of the system.  Taken
  together, these results suggest that this young galaxy is undergoing
  its first major epoch of mass assembly.
\end{abstract}


\section{Introduction}

Deep imaging surveys, particularly with {\it Hubble Space Telescope
  (HST)} have now uncovered thousands of star-forming galaxies in the
redshift range $z=4-6$
(eg. \citealt{Ouchi04,Giavalisco04,Yoshida06,Beckwith06,Bouwens08}).
Studying such galaxies are key steps towards understanding the physical
processes that drive galaxy formation in the early Universe, probing
the properties of galaxies when they formed their first generation of
stars.  Concentrated follow-up of Lyman-break galaxies at this epoch
has provided estimates of stellar masses
($\sim$10$^{9-10}$M$_{\odot}$), ages ($\sim$150-300\,Myr) and hence
inferred star-formation rates of a few 10's M$_{\odot}$\,yr$^{-1}$
\citep{Verma07,Stark09,McLure09}.  With high specific star-formation
rates, these young galaxies are likely assembling their first
significant stellar mass.  Moreover, since this epoch is within
$\sim$300\,Myr of the end of reionisation (assuming reionisation was
completed by $z\sim$6) many of these galaxies should have been actively
star-forming at $z\gsim$7.  As such, probing the ubiquity of
star-forming galaxies, the ionising photon density from star-formation
and their relation to AGN are important issues, potentially allowing us
to understand how and when the first galaxies were assembled and how
the Universe was reionised.  Moreover, theoretical models suggest that
the most rapid epoch of dark halo assembly for galaxies such as the
Milky-Way occur at $z\sim$5 \citep{MoWhite02,Okamoto05}.  Through
detailed studies of the physical properties of galaxies at this epoch
(such as masses and star-formation rates), we can accurately assess
their contribution to the mass growth of galaxies like the Milky-Way,
and their effect on the gaseous IGM.

Indeed, one of the key findings in recent years is the discovery that a
significant fraction of high-redshift galaxies are surrounded by
``superwinds''
(e.g.\ \citealt{Franx97,Pettini02a,Shapley03,Bower04,Wilman05,Swinbank07a})
-- starburst and/or AGN driven outflows which expel gas from the galaxy
potential and hence play no further role in the star-formation history
of the galaxy.  Such feedback processes may offer natural explanations
to the shape and normalisation of the local luminosity function
\citep{Benson03,Bower06,Croton06} and offer natural explanations as to
why only 10\% of the baryons cool to form stars
\citep{White78,Balogh01_ccc}.  Although velocity offsets between
star-forming regions and outflowing gas of several hundred km/s have
been measured (suggestive of large scale outflows), constraining the
geometry and mass loading is vital to test whether the outflows are
truly large-scale (as often observed in low-redshift Ultra-Luminous
Infrared Galaxies (ULIRGS); \citealt{Martin05}), or whether the
outflows are confined to individual star-forming regions.  Such
constraints are vital in order to test whether the outflowing material
escapes into the IGM, or whether it eventually stalls, fragments and
drains back down onto the galaxy disk, causing further bursts of
star-formation.

However, probing the masses, star-formation rates, and interaction
between star-formation and gas dynamics which are routine at $z\sim$2
(eg. \citealt{Genzel06,ForsterSchreiber06,Swinbank06b,Law07,Stark08,Lehnert09,Law09})
are difficult beyond $z\sim$3 due to a combination of surface
brightness dimming and the smaller physical sizes of the earliest
systems (and the strong evolution in the UV luminosity function between
$z\sim$3--6 means there are fewer luminous sources;
eg. \citealt{Stark09}).  Prior to the launch of {\it James Webb Space
  Telescope (JWST)} one particularly appealing route to study the
internal structures of primitive galaxies at $z\sim5$ is to target
galaxies which have been highly magnified by their serendipitous
alignment with a foreground massive cluster.  The lensing magnification
(up to a factor 30$\times$ in the most extreme cases) has two effects:
the galaxy image is amplified at a fixed surface brightness so that the
galaxy image appears both brighter and larger.  By combining
gravitational lensing with resolved spectroscopic imaging, detailed
maps of the sizes, dynamics and hence masses of both young galaxies and
even starburst/H{\sc ii} regions within these galaxies can be made.
Such observations can be used to investigate whether the physical
conditions for star-formation were significantly different from
comparably luminous galaxies at $z\sim2$ and galaxies today.  Indeed,
gravitational lensing and resolved spectroscopic imaging has been used
to investigate galaxy dynamics on $\sim$1\,kpc scales from $z=1$--3
(eg.\ \citealt{Swinbank03,Swinbank06a,Salucci07,Nesvadba06}), whilst
coupling lensing with adaptive optics (AO) source-plane studies can
even reach $\lsim$100\,pc (\citealt{Stark08}; Jones et al.\ 2009 in
prep).

Studying the physical processes of galaxy formation on scales
comparable to the largest H{\sc ii} regions is likely to have broad
implications for our understanding of the dominant mechanism by which
galaxies at high redshift assemble their stellar mass, and what
processes drive the star-formation activity \citep{Bournaud09}.  For
example, recent hydro-dynamical simulations have suggested that
continuous gas accretion (so-called ``cold-flows'') may play an
important role in driving star-formation at $z>2$
\citep{Dave08,Dekel09}.  In such models, these cold streams keep the
rotating disk configuration intact, but with turbulent star-forming
clumps which eventually merge to form a central spheroid.  This is a
provocative result, as it has generally been accepted that major
mergers are the dominant mechanism by which massive galaxies assemble
their stellar mass \citep{Conselice08}.  However, to test such models
in detail, the galaxy velocity field and the turbulent speeds on scales
comparable to individual H{\sc ii} regions in a large sample of
high-redshift, star-forming galaxies are required.  Resolved dynamics
of galaxies at $z\sim2$ have lent support to the cold-stream model, but
at the spatial scales available in non-lensed galaxies (1--4\,kpc) it
is difficult to draw definitive conclusions.  Nevertheless, the
advances in instrumentation (particularly sensitive integral-field
units and AO) are beginning to allow studies of the kinematics,
distribution of star-formation, gas and stars and chemical properties
in high-redshift star-forming galaxies which can be fed into (and thus
test) numerical simulations.

In this paper we report the detailed follow-up of a lensed galaxy at
$z=4.92$ behind the lensing cluster MS\,1358+62, first studied in
detail by \citet{Franx97}.  We use the {\it HST} ACS \& NICMOS imaging
to investigate the rest-frame UV/optical morphology.  Together with
{\it Spitzer} IRAC imaging we construct the spectral energy
distribution (SED) to estimate the stellar mass.  Using the Gemini/NIFS
integral field unit we map the [O{\sc ii}]$\lambda\lambda$3726.2,3728.9
emission line strength and dynamics, distribution of star-formation and
outflow energetics.  We have adopted a standard $\Lambda$CDM cosmology
with $H_{0}=72\kms$, $\Omega_{M}=0.27$ and $\Omega_{\Lambda}=0.73$.
All quoted magnitudes are on the AB system.

\begin{figure*}
\centerline{
  \psfig{figure=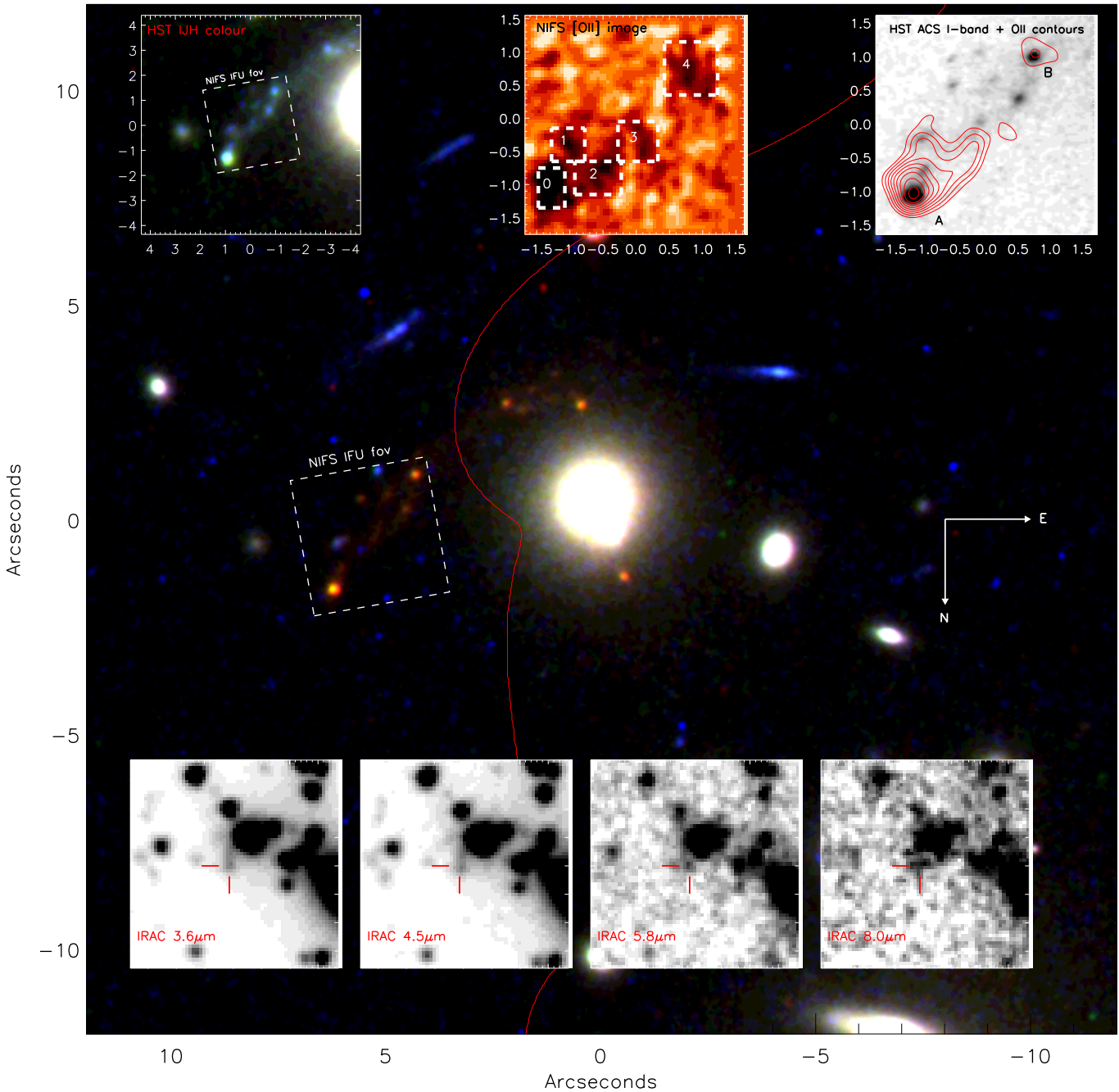,width=7in,angle=0}}
\caption{True colour {\it HST} ACS $VRI$-band image of the lensed
  $z=4.92$ galaxy behind MS\,1358+62.  The image is centered at the
  position of a bright cluster elliptical at $\alpha=$13:39:49.51
  $\delta=$+62:30:48.69 (note that the BCG lies approximately 20$''$
  south).  The $z=4.92$ critical curve from our best-fit lens model is
  also overlaid.  The {\it HST} imaging clearly shows the galaxy
  resolved into two images of the background galaxy, with the
  morphology dominated by up to six star-forming regions surrounded by
  a diffuse halo.  The dashed white box shows the NIFS IFU field of
  view used to cover the $z=4.92$ galaxy image.  The upper left insets
  show a colour image generated from the NICMOS F110 and F160 images
  showing that the $J$- and $H$-band morphologies clearly trace the
  observed $VRI$-band morphology.  The upper middle panel shows the
  white light (wavelength collapsed) image of the [O{\sc ii}] emission
  from the IFU datacube (dark grey denotes highest intensity).  The
  boxes (labeled 0-4) denote the regions from which spectra were
  extracted (see Fig.~4).  The upper right insets show the {\it HST}
  ACS $I$-band image (dark grey denotes high intensity) with contours
  from the [O{\sc ii}] emission line image overlaid, showing that the
  [O{\sc ii}] emission line morphology follows that seen in the
  rest-frame UV.  The contours denote [O{\sc ii}] emission line flux
  levels of
  0.2--1$\times$10$^{-16}$\,erg\,s$^{-1}$\,cm$^{-2}$/sq-arcsec in
  increasing in units of
  0.1$\times$10$^{-16}$\,erg\,s$^{-1}$\,cm$^{-2}$/sq-arcsec.  The lower
  four insets show the {\it Spitzer} IRAC 3.6, 4.5, 5.8 and 8.0$\mu$m
  images around the lensed galaxy.  In each panel we mark the position
  of the brightest component of the $z=4.92$ arc, which is clearly
  visible in all four channels (each of these panels is 30$"$ on a
  side). }
\label{fig:HSTcol}
\end{figure*}

\section{Observations and Data Reduction}
\label{sec:obs_red}

\subsection{Optical-, near- and mid-infrared imaging}

Details of the optical, near- and mid- infrared observations of this
galaxy cluster are given in \citet{Richard08}.  Briefly, the lensing
cluster MS\,1358+62 (z$_{cl}$=0.33; $\alpha$: 13:59:50 $\delta$:
+62:31:05, J2000) was observed with {\it HST} ACS and NICMOS as part of
program PID: 9717 \& 10504 using the 435W, 475W, 625W, 775W, 850LP,
F110W and F160W filters with integration times 5.4-16\,ks each in the
optical bands and $\sim$19\,ks each in the near-infrared bands.  The
optical and near-infrared imaging reach typical 5$\sigma$ depths of
AB=26.5--27.5 magnitudes.  We also include $K$-band photometry from
\citet{Soifer98} which reaches a 5$\sigma$ sensitivity of
$K_{AB}\sim$23.  The {\it Spitzer} IRAC 3.6, 4.5, 5.8 and 8.0$\mu$m
imaging comprises 2.4\,ks in each band, reaching 5$\sigma$ depths of
AB$\sim$24 \citep{Richard08}.

The $z=4.92$ arc lies approximately 21$''$ south west of the brightest
cluster galaxy at $\alpha$: 13:59:48.7 $\delta$: +62:30:48.34 (J2000)
and was discovered during a spectroscopic survey of this cluster by
\citet{Franx97} in which the redshifted Ly$\alpha$ and UV-ISM lines
were used to derive an unambiguous redshift of $z=4.92$ (a companion
galaxy at the same redshift located approximately 200\,kpc away in the
source-plane was also discovered during the same survey).  In
Fig.~\ref{fig:HSTcol} we show a true-colour {\it HST} ACS $VIz-$band
image of the cluster core around the $z=4.92$ arc.  We also show an
ACS/NICMOS $IJH$-band true-colour image, as well as thumb-nails around
the galaxy in the four IRAC images showing that the galaxy is detected
from $V$-band to the IRAC 8.0$\mu$m channel.

To calculate the photometry of the lensed galaxy we first consider the
bright elliptical galaxy locate $\sim$6$''$ due East (which has a low
surface brightness halo which extends beyond the $z=4.92$ galaxy
image).  We construct a model of the elliptical galaxy surface
photometry using the {\sc iraf bmodel} ellipse fitting algorithm
\citep{Jedrzejweski87} (we note that we mask the $z=4.92$ galaxy image
during the fit).  In this model, the surface photometry includes the
$c_4$ Fourier coefficient (which is necessary to describe the boxiness
of the isophotes), but all other Fourier terms were forced to zero.  We
then subtract the best fit model from the galaxy image and then use
{\sc sextractor} \citep{Bertin96} to estimate the residual background
within the frame.  Since accurate colours are required to derive the
spectral energy distribution we calculate the magnitude of the lensed
galaxy in the various pass bands using an elliptical aperture
(approximately 3$''\times$1.5$''$ centered on the first image of the
arc as covered by the NIFS IFU).  This aperture is constructed from the
coadded $I$- and $z$-band images since these represent the highest
signal-to-noise in the z=4.92 arc.  The same aperture is then applied
to all of the images to measure the photometry.  In the IRAC channels
we apply an aperture correction to account for the flux lying outside
the aperture ($\sim$0.6-0.8 magnitudes from 3.6--8.0$\mu$m) as
determined from bright (unsaturated) point sources in the image.

\begin{figure}
\centerline{
  \psfig{figure=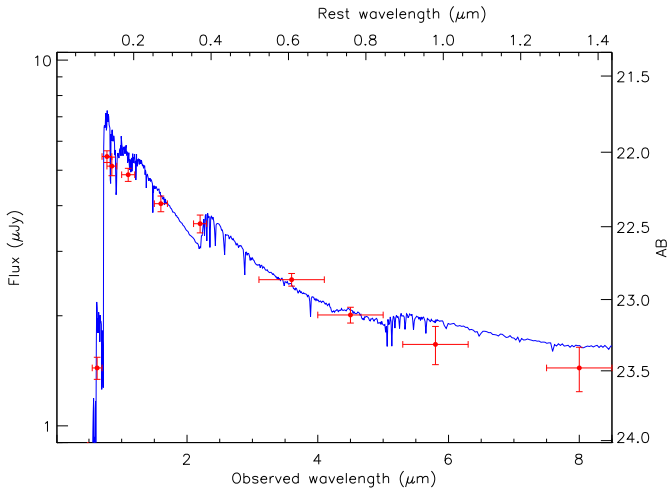,width=3.5in,angle=90}}
\caption{Observed Spectral Energy Distribution (SED) of the galaxy from
  the observed {\it HST} ACS/NICMOS and {\it Spitzer} IRAC photometry.
  We overlay the best-fit SED from {\sc hyperz} which predicts a young
  age of $\sim$20\,Myr and moderate (A$_{v}$=0.25) reddening.  Fitting
  the SED using we derive a stellar mass (corrected for lensing
  amplification) of $M_{\star}=7\pm2\times10^{8}$M$_{\odot}$.  }
\label{fig:SED}
\end{figure}

In Table~1 we give the observed photometry of the galaxy from the {\it
  HST} ACS \& NICMOS and {\it Spitzer} IRAC imaging and in
Fig.~\ref{fig:SED} we show the observed photometry with the best fit
model SED overlaid (see \S~\ref{sec:stellmass}).  We note that the
large error-bars (particularly at 5.8 and 8.0$\mu$m) reflect the
uncertainty in recovering the photometry from the arc after modeling
and removing the foreground elliptical galaxy which dominates the
light, particularly at the reddest wavelengths.  We also note that
there is MIPS 24$\mu$m coverage of this cluster with a bright source at
the position of the galaxy.  However, due to the blending with the
foreground lensing galaxy (located $\sim$6$''$ East), it is not
possible to reliably deblend the photometry from the elliptical galaxy.

\subsection{Spectroscopic Imaging}

\begin{figure}
\centerline{
  \psfig{figure=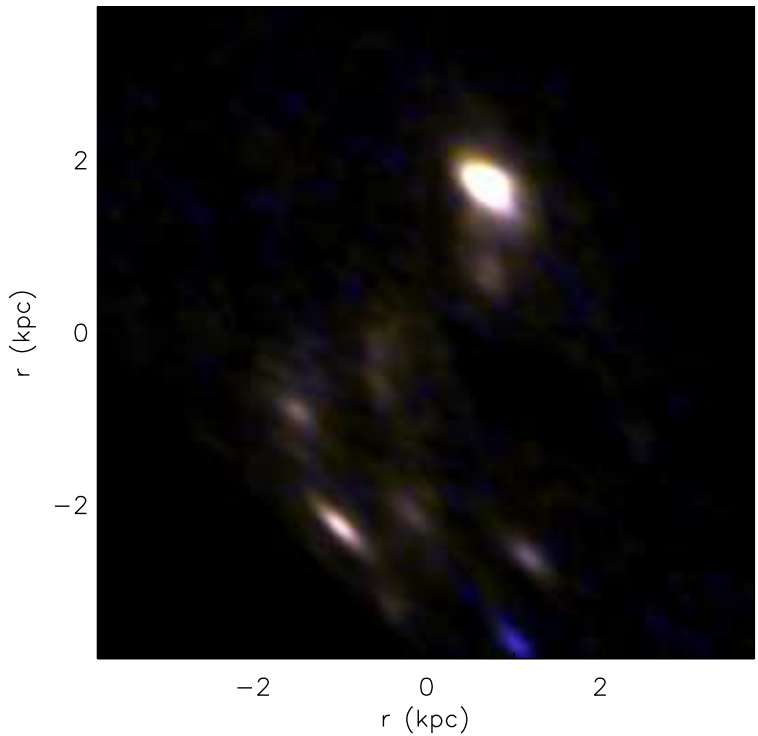,width=4in,angle=90}}
\caption{True colour {\it HST} ACS $VRI$-band reconstruction of the
  lensed galaxy using the mass model described in
  \S\,\ref{sec:analysis}.  The amplification of the galaxy is a factor
  12.5$\pm$2.0$\times$.  In the source plane, the galaxy has a spatial
  extent of $\sim$4\,kpc and comprises at least five discrete
  star-forming clumps.  The largest of these is only marginally
  resolved, with a rest-frame optical half light radii of
  $\sim$200\,pc.}
\label{fig:HSTcol_source}
\end{figure}

Three dimensional spectroscopic imaging observations around the
redshifted [O{\sc ii}]$\lambda\lambda$3726.2,3728.9\AA\ emission line
were taken with the Gemini-North Near-Infrared Integral Field
Spectrometer (NIFS) between 2006 February 08 and 2006 February 09
during science verification time.

The Gemini-NIFS IFU uses an image slicer to take a 3.0$''\times$3.0$''$
field and divides it into 29 slices of width 0.103$''$. The dispersed
spectra from the slices are reformatted on the detector to provide
2-dimensional spectro-imaging, in our case using the $K$-band grism
covering a wavelength range of 2.00--2.43\,$\mu$m.  The observations
were performed using an ABC configuration in which we chopped by 6$''$
to blank sky to achieve sky subtraction. Individual exposures were
600\,s and each observing block was 3.6\,ks, which was repeated four
times for resulting in an integration time of 14.4\,ks (with a total of
9.6\,ks on source and 4.8\,ks on sky).

We reduced the data with the standard Gemini {\sc iraf nifs} pipeline
which includes extraction, sky subtraction, wavelength calibration, and
flat-fielding.  Residual OH sky emission lines were removed using the
sky-subtraction techniques described in \citet{Davies07}.  To
accurately align and mosaic the individual datacubes we created white
light (wavelength collapsed) images around the redshifted [O{\sc ii}]
line from each observing block and centroid the galaxy within the data
cube.  These were then spatially aligned and co-added using an average
with a 3$\sigma$ clipping threshold to remove remaining cosmetic
defects and cosmic rays.  Flux calibration was carried out by observing
a bright A0V standard star (HIP\,78017) at similar airmass to the
target galaxy immediately after each observing block.  From the reduced
standard star cube, we measure a $K$-band seeing of FWHM=0.45$''$.  We
note that accounting for lensing amplification factor
12.5$\pm$2.0$\times$ (derived in \S\ref{sec:analysis}) this corresponds
to a source-plane FWHM of $\sim$0.6\,kpc.  
The spectral resolution of the data (measured from the sky-lines at
$\sim$2.2$\mu$m) is $R=\lambda/\Delta\lambda$=5300 which corresponds to
$\sigma$=3\AA\ or 25\,km\,s$^{-1}$. In all following sections, quoted
line widths are deconvolved for the instrumental resolution.

\section{Analysis}
\label{sec:analysis}

\subsection{Lens modeling}

We first estimate the amplification of the galaxy by making use of the
updated gravitational lens model from \citet{Richard08}.
Fig.~\ref{fig:HSTcol} shows the cluster region around the lensed
galaxy with the best-fit $z=4.92$ critical curve overlaid.  Using the
mapping between image and source-plane co-ordinates, we ray-trace the
$z=4.92$ galaxy image and reconstruct the source-plane morphology,
deriving a luminosity weighted magnification factor
$\mu$=12.5$\pm$2.0$\times$ (which corresponds to $\Delta$m=2.7$\pm$0.2
magnitudes).  The error-bar on $\mu$ is derived by ray-tracing the
family of acceptable gravitational lens models which adequately
reproduce the multiply-imaged galaxies in the cluster.

Accounting for lensing amplification, the intrinsic magnitude of the
galaxy is $I_{AB}=24.94\pm0.08$.  For comparison, deep imaging surveys
of Lyman-break galaxies at $z\sim$5 have derived an $L^{*}$ of
$i_{AB}\sim25.3$ suggesting that the $z=4.92$ galaxy studied here is
typical of the UV-continuum selected samples at this redshift
(eg. \citealt{Ouchi04}).

In Fig.~\ref{fig:HSTcol_source} we show the reconstructed source-plane
image of the galaxy which shows that the rest-frame UV/optical
morphology is elongated, but is dominated by two star-forming regions
separated by $\sim$4\,kpc with at least four more lower-surface
brightness regions within the galaxy and surrounded by a low surface
brightness halo.

To estimate the size of the source-plane PSF in each band, we first
measure the size of the PSF using the $R\sim$19 star located 5$''$ due
south of the arc, which has a FWHM=0.10, 0.33, 0.30$''$ in the observed
$I$-, $J$- and $H$-bands respectively.  The two largest star-forming
regions (at approximately [1.8,2.0] and [-1.0,-2.1]\,kpc in
Fig.~\ref{fig:HSTcol_source}) have an $I$-band FWHM which is marginally
resolved (FWHM=0.17$\pm$0.03$''$), but both are unresolved in $J$ and
$H$-bands.  Since the lensing amplification of the source has a
preferential direction, we reconstruct the source-plane image of the
star at the locations of the brightest star-forming regions and find
that the ACS PSF transforms to 0.06$''\times$0.014$''$,
(320$\times$85\,pc). Both of the brightest two H{\sc ii} regions are
marginally resolved in the $I$-band imaging (FWHM=0.17$\pm$0.03$''$ and
FWHM=0.15$\pm$0.04 for regions $A$ and $B$ in Fig.~\ref{fig:HSTcol}
respectively) corresponding to FWHM=500$\pm$100\,pc and
FWHM=400$\pm$150\,pc (deconvolved for PSF).  We note that the size of
the largest of these two H{\sc ii} regions is comparable to the size
measured by \citet{Franx97} (FWHM$\sim$390\,pc) from {\it HST} WFPC2
observations.  In the $J$- and $H$-bands, the source-plane PSFs is
approximately 800$\times$220\,pc and 300$\times$1\,kpc in the NIFS
[O{\sc ii}] emission line map.  We therefore conservatively suggests
sizes of FWHM$\lsim$500 and 400\,pc in $J$- and $H$-bands respectively
and $\lsim$600\,pc in the O{\sc ii} emission line map for the two H{\sc
  ii} regions, $A$ \& $B$ respectively.  We discuss the implications of
these sizes in \S3.6, but note that clearly higher spatial resolution
IFU observations using laser-guide star AO would be very valuable to
further constrain their sizes.

\begin{figure}
\centerline{
  \psfig{figure=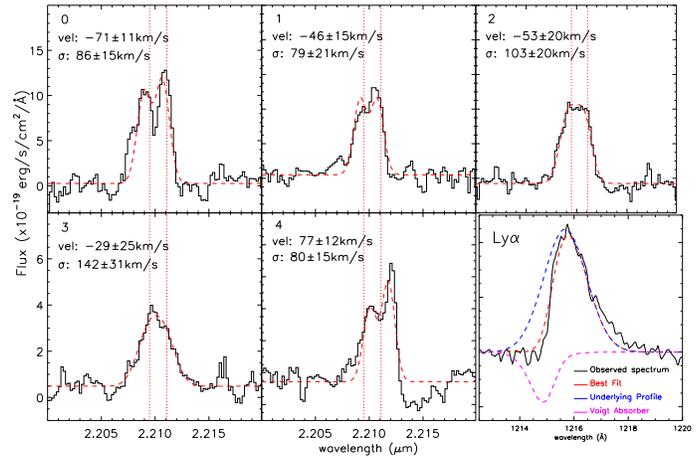,width=3.5in,angle=0}}
\caption{One-dimensional spectrum of the five star-forming regions
  within the $z=4.92$ galaxy from the NIFS IFU observations.  In all
  panels we show the position of the [O{\sc
      ii}]$\lambda\lambda$3726.8,3728.9 doublet at a fixed redshift of
  $z=$4.9296.  In each panel we also overlay the best fitting spectrum
  and give the (rest-frame) velocity offset from the systemic redshift
  and line width of the best fitting double Gaussian profile.  From the
  spectra, the velocity gradient of $\Delta v$=180$\pm$20\,km\,s$^{-1}$
  can easily be seen.  The final panel shows the one-dimensional
  spectrum of the $z=4.92$ galaxy around the Ly$\alpha$ emission line
  (from \citealt{Franx97}) showing the asymmetric line profile.  Using
  the systemic redshift measured from the nebular emission, we
  de-redshifted the spectrum and find that the asymmetric emission line
  is best fit with an underlying Gaussian emission line profile (blue
  dashed line) combined with a Voigt absorber in the blue wing.}
\label{fig:OIIspec}
\end{figure}

\subsection{Stellar mass}
\label{sec:stellmass}

\citet{Soifer98} use optical and near-infrared photometry to estimate
the stellar mass in this galaxy, deriving
M$_{\star}$=2--16$\times$10$^{9}$M$_{\odot}$.  Here, we make use of the
much deeper ACS/NICMOS and, particularly IRAC imaging in order to
improve the estimate of the stellar mass.  We infer the stellar mass of
the galaxy by fitting the latest Charlot \& Bruzual (2007) (hereafter
CB07) stellar population synthesis models to the observed SED (which
include an improved treatment of the thermally-pulsating asymptotic
giant branch (TP-AGB) phase of stellar evolution).  Following
\citet{Stark09}, we allow 200 age steps from 10$^{5}$ to 10$^{9}$\,yr
(approximately logarithmically spaced), with a \citet{Salpeter55}
initial mass function (IMF) and consider both a solar and sub-solar
(0.2\,Z$_{\odot}$) metallicity.  We fit both constant star-formation
histories and exponentially decaying ($\tau$) models with $e$-folding
decay times of 100\,Myr.  We consider the effects of dust in the
modeling by adopting the \citet{Calzetti94} reddening law (using colour
excesses of $E(B-V)$=0.0, 0.03, 0.05, 0.1, 0.3 \& 1.0).  The resulting
template SEDs are convolved with the filter response curves, and by
comparing the observed magnitudes and their associated errors, we
compute a reduced $\chi^2$ value for each set of parameters (age,
$E(B-V)$ and normalistation).  The 1$\sigma$ uncertainties are
calculated from the parameter range which produce $\Delta\chi^2$=1 with
respect to the best fit SED (whilst marginalising over the other free
parameters).

For consistency, we first model the $IJHK$-band galaxy photometry from
\citet{Soifer98}, deriving acceptable solutions with
$E(B-V)$=0.06--0.4, ages=1$\times$10$^{6-9}$\,Myr and stellar masses of
M$_{\star}$=1$\times$10$^{9}$--4$\times$10$^{10}$\,M$_{\odot}$ (which
is consistent with \citealt{Soifer98}).  However, the much deeper {\it
  HST} ACS/NICMOS and {\it Spitzer} photometry ($VIzJHK$ \&
IRAC\,3.6,4.5,5.8,8.0$\mu$m) allows much better constraints to be
derived, and we infer a best fit SED with an age of 14$\pm$7\,Myr, an
$E(B-V)$=0.05$\pm$0.05 and a stellar mass of
$M_{\star}$=7$\pm$2$\times$10$^{8}$\,M$_{\odot}$, with a
0.2\,Z$_{\odot}$ metallicity template marginally prefered over solar
metallicity (although formally the difference in $\Delta\chi^2$ is less
than 1\,$\sigma$ between the two).  We note that the constant
star-formation and $\tau$ models provide equally good fits to the data,
with stellar masses which vary by $<$5\%.  We also attempt fits
excluding the IRAC\,5.8 \& 8.0$\mu$m photometry, but the resulting
stellar mass and age agree within the 1$\sigma$ errors.  We note that
the stellar mass derived here is significantly lower (at least a factor
$>$3$\times$) than that derived by \citealt{Soifer98}, which most
likely arises due to the improved photometric constraints and very
young age as suggested by the IRAC photometry.

\subsection{Star-formation rate and dynamics}

Next we use the spatially resolved [O{\sc
    ii}]$\lambda\lambda$3726.8,3728.9 emission line maps from our NIFS
IFU observation to estimate the distribution of star-formation and
dynamics within the galaxy.  As Fig.~\ref{fig:HSTcol} shows, the
nebular line emission is clearly patchy, and dominated by the brightest
regions seen in the $HST$ ACS and NICMOS imaging.  Collapsing the
brightest regions within the galaxy we find that the integrated [O{\sc
    ii}] emission line flux is
$f_{[OII]}$=1.5$\pm$0.3$\times$10$^{-16}\cgs$.  Accounting for an
amplification factor ($\mu$=12.5$\pm$2.0$\times$) and using the
conversion between [O{\sc ii}] emission line flux and star-formation
rate from \citet{Kennicutt98} we derive an intrinsic star-formation
rate of $SFR_{[OII]}$=42$\pm$8\,M$_{\odot}$\,yr$^{-1}$ (uncorrected for
reddening).  This star-formation rate is consistent with the 3$\sigma$
upper-limit of SFR$\sim$80\,M$_{\odot}$\,yr$^{-1}$ (corrected for
lensing) from SCUBA 850$\mu$m imaging \citep{Knudsen08}.  We note that
if the star-formation rate derived from the [O{\sc ii}] emission line
flux ($SFR$=42$\pm$8\,M$_{\odot}$\,yr$^{-1}$) has been sustained, then
it takes just $\sim$15\,Myr to build a stellar mass of
M$_{\star}$=7$\times$10$^{8}$\,M$_{\odot}$, suggesting the current
burst may be the first major episode of star-formation within this
galaxy (which is also consistent with the young SED).  It is also
useful to note that the starburst intensity (SFR/area) for the galaxy
(S$_e\sim$4$\times$10$^{11}$\,L$_{\odot}$\,kpc$^{-2}$) is comparable
with local- (eg. \citealt{Meurer97}) and high redshift-
(eg. \citealt{Giavalisco96,Steidel96}) galaxies, suggesting the same
mechanisms may limit the starburst intensity at $z\sim$5 as those at
lower redshifts.

We also use the spatially resolved [O{\sc ii}] emission line doublet
from the NIFS IFU observations to investigate the dynamics of the
galaxy.  By collapsing the velocity field from the brightest regions
within the datacube we derive a peak-to-peak velocity gradient of
$\Delta v$=180$\pm$20\,km\,s$^{-1}$ across 4\,kpc in projection
(Fig.~\ref{fig:vrot}).  If these dynamics reflect virialised motion
within a bound system, then we estimate a dynamical mass of
M$_{dyn}$($<$2\,kpc)=3$\pm$1$\times$10$^{9}${\it
  csc$^2$(i)}\,M$_{\odot}$.  Adopting a canonical inclination of
30$^{\circ}$ suggests a dynamical mass of
M$_{dyn}\sim$10$^{10}$M$_{\odot}$.  Alternatively, the dynamical mass
can also be estimated via the velocity dispersion using the relation
M$_{dyn}$=$C\sigma^2 r/G$.  The factor $C$ depends on the galaxy mass
distribution and velocity field, and ranges from C$\lsim$1 to $C\gsim$5
(depending on the mass density profile, velocity anisotropy and
relative contribution to $\sigma$ from random motion, rotation and
assumption of spherical or disk-like system; see \citet{Erb06a} for a
detailed discussion).  Adopting $C$=3.4 (such that a direct comparison
to star-forming galaxies at $z$=2--3 can be made; \citealt{Erb06a}), we
derive a dynamical mass of M$_{dyn}\sim10^{10}$\,M$_{\odot}$.  This
dynamical mass is approximately 6-10$\times$ smaller than the median
Lyman-break galaxy mass at $z\sim$3 for which masses have been measured
using similar techniques
(eg.\ \citealt{Erb06a,ForsterSchreiber06,Law09}), but comparable to
that of the only other star-forming galaxy at these redshifts with
measured nebular emission line dynamics (RCS0224-002\,arc at z=4.88;
\citealt{Swinbank07a}).  We also note that the inferred dynamical mass
is much larger than the stellar mass.  If the central regions are
baryon dominated, this may suggest that the gas reservoir makes up
$>$75--90\% of the baryons.  Clearly observations of the cold molecular
gas (such as the redshifted CO(1-0) or CO(3-2)) would be required to
test this.

Next, we investigate the properties of individual star-forming regions
within the galaxy.  In Fig.~\ref{fig:OIIspec} we show the extracted the
spectra from the strongest line emitting regions from the galaxy.  The
two largest star-forming regions in Fig~\ref{fig:HSTcol} have velocity
dispersions of $\sigma_0$=86$\pm$15 and
$\sigma_4$=80$\pm$15\,km\,s$^{-1}$.  Taken with their sizes derived
above we derive masses of star-forming regions of order
6--9$\times$10$^{8}$\,M$_{\odot}$ (we note that in calculating these
masses, we have used the same value of $C$ as above for simplicity),
and star-formation rates of SFR$_0$=12$\pm$1 and
SFR$_4$=6$\pm$1\,M$_{\odot}$yr$^{-1}$ respectively suggesting that up
to half of the total star-formation is occurring within two dense
regions and outside the nuclear regions.

\begin{center}
\small \centerline{\sc Table 1.}  \centerline{\sc Observed Aperture Photometry for MS\,1358+62arc}
\medskip
\begin{tabular}{lccc}
\hline\hline
\noalign{\smallskip}
Filter & mag (AB)\\
\hline
\noalign{\smallskip}
$B_{475}$     & $>$26.5   \\
$V_{625}$     & 23.50$\pm$0.08  \\
$I_{775}$     & 22.05$\pm$0.05  \\
$z_{850}$     & 22.12$\pm$0.05  \\
$J_{110}$     & 22.18$\pm$0.08  \\
$H_{160}$     & 22.38$\pm$0.08  \\
$K$          & 22.52$\pm$0.20 \\
3.6\,$\mu$m  & 22.90$\pm$0.10 \\
4.5\,$\mu$m  & 23.14$\pm$0.10 \\
5.8\,$\mu$m  & 23.43$\pm$0.25 \\
8.0\,$\mu$m  & 23.50$\pm$0.25 \\
\noalign{\smallskip}
\hline\hline
\label{tab:photom}
\end{tabular}
\end{center}
\medskip
\begin{minipage}{3.5in}
  \small {\sc Note.} -- To convert to intrinsic magnitudes (corrected
  for the lensing amplification factor $12.5\pm{2.0}\times$) add
  2.7\,mags.
\end{minipage}

\subsection{Comparison with rest-frame UV}

\citet{Franx97} discuss the morphology and rest-frame UV spectral
properties of this galaxy in detail.  In particular the UV-ISM
absorption lines show velocity variations on the order of
200\,km\,s$^{-1}$ along the arc with the Si{\sc ii}$\lambda$1260 line
systematically blue-shifted with respect to the Ly$\alpha$ emission, and
an asymmetric Ly$\alpha$ emission line with a red tail.  As discussed
by \citet{Franx97} these spectral features are naturally explained by
an outflow model, in which the blue side of the Ly$\alpha$ line has
been absorbed by outflowing neutral H{\sc i}.  The description of the
line profile is very typical of the emission profiles seen in other
Lyman-break galaxies.  Indeed, velocity offsets between the nebular
emission lines (such as H$\alpha$) and UV-ISM emission/absorption lines
(such as Ly$\alpha$ and Si{\sc ii}$\lambda$1260) are now common in
high-redshift star-forming galaxies, and are usually interpreted as
evidence for large-scale starburst driven outflow
\citep{Erb03,Teplitz00,Pettini02a}.  Such a model provides a good
description of the integrated properties of high-redshift galaxies and
more detailed observations of local starbursts
(eg.\ \citealt{Pettini02a,Shapley03,Tenorio-Tagle99,Heckman00,Grimes06}).
In this model, the Ly$\alpha$ emission comes from photons emitted from
the star-forming regions.  To reach the observer these must pass
through part of the foreground (blue shifted) shell. This absorption
causes the peak emission wavelength to appear redshifted relative to
the nebular emission lines.  In addition, photons may be scattered or
emitted from the receding shell.  Photons that are either created on
the inner surface of the shell (e.g. by UV irradiation from the
starburst), or multiply scattered within the receding shell will
acquire the mean velocity of the outflow and will be seen as redshifted
by the observer (see the discussion of \citealt{Hansen06}).

\begin{figure}
\centerline{ \psfig{figure=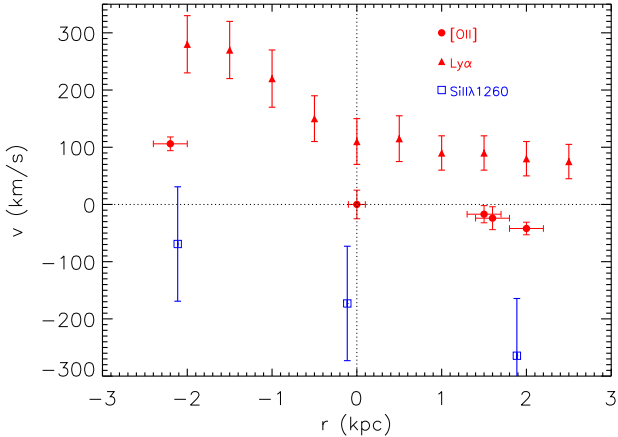,width=3.5in,angle=90}}
\caption{Extracted, one-dimensional velocity gradient along the long
  axis of the galaxy (source plane).  The velocity gradient of
  $\sim$180\,km\,$s^{-1}$ observed in the [O{\sc ii}] emission is
  mirrored by the Ly$\alpha$ and UV ISM lines, although they appear
  systematically offset from the systemic by 150--200\,km\,s$^{-1}$,
  suggesting that the galaxy is surrounded by a young galactic scale
  outflow which has yet to decouple and burst out of the galaxy.  Using
  the velocity offsets and spatial extent of the galaxy, we suggest the
  outflow is $\lsim$15\,Myr old.}
\label{fig:vrot}
\end{figure}

With the systemic velocity of the galaxy measured from the [O{\sc ii}]
emission, we can compare the nebular emission line velocity gradient
seen in this galaxy with those observed in Ly$\alpha$ emission and
UV-ISM lines, as well as crudely estimating the energetics of the
outflow.  Using the systemic redshift measured from the [O{\sc ii}]
emission line doublet ($z=$4.9296$\pm$0.0002) the Ly$\alpha$ appears
systematically redshifted from the systemic redshift by
100--200\,km\,s$^{-1}$ across the 4\,kpc extent of the galaxy
(Fig.~\ref{fig:vrot}).  Moreover, the UV-ISM absorption lines are also
correspondingly blue shifted by $\sim$200$\pm$100\,km\,s$^{-1}$.  This
is typical of high-redshift, star-forming galaxies.  However,
strikingly, the velocity gradient in the nebular emission is mirrored
in both the Ly$\alpha$ and UV-ISM lines.  The fact that the velocity
structure of the outflow follows that seen in the nebular emission
suggests the outflow has yet to decouple from the galaxy.  Indeed,
assuming that the outflow is no larger than the spatial extent of the
stellar component of the galaxy ($\lsim$2\,kpc), then for a constant
velocity of 150\,km\,s$^{-1}$ it takes just 15\,Myr to travel 2\,kpc.
This suggests that the young outflows from individual H{\sc ii} regions
have not yet merged to form a superwind surrounding the galaxy.

To test whether a young outflow is energetically feasibly, first we
de-redshift the rest-frame UV spectrum using the systemic redshift.  In
Fig.~\ref{fig:OIIspec} we show the region around the Ly$\alpha$
emission in the rest-frame, clearly showing the asymmetric profile.  A
single Gaussian profile fit suggests a velocity offset from the
systemic redshift of $\sim$200\,km\,s$^{-1}$.  However, a much better
fit ($\Delta\chi^2>$25) is obtained by modeling the emission line with
an underlying Gaussian profile (fixed in wavelength at 1215.67\AA\,)
with an Voigt profile absorber in the blue wing.  In this fit, the
emission line width and intensity of the underlying Gaussian profile
are allowed to vary, as are the velocity, impact parameter ($b$) and
column density ($n$) of the Voigt profile.  The best fit model has a
Voigt profile centered -300$\pm$100\,km\,s$^{-1}$ from the systemic
with $b\sim70$\,km\,s$^{-1}$ and $n\sim4\times$10$^{17}$\,cm$^{-2}$.
Allowing the centroid of the underlying Gaussian emission line profile
to vary, the column density increases to
$n\sim3\times$10$^{18}$\,cm$^{-2}$ and $b\sim$120\,km\,s$^{-1}$ with an
underlying Gaussian profile with centroid $\Delta
v=-100\pm30$\,km\,s$^{-1}$ from that predicted from the [O{\sc ii}]
emission (we note that the improvement in $\chi^{2}$ between the two
fits is $\Delta \chi^{2}<4$ and so indistinguishable, although fixing
the underlying Gaussian profile at the systemic redshift is more
physical).  We caution that both of these column densities are near the
flat part of the curve of growth causing degeneracies between N$_{H}$
and $b$ and so come with considerable uncertainty.  However, we can at
least test whether such a value is reasonable using the UV absorption
lines and assuming an ionisation state for the ISM.  Using the
rest-frame UV spectroscopy we measure (rest-frame) equivalent widths of
$W_{o}$(Si{\sc ii}$\lambda$1260.4)=0.30$\pm$0.07\AA\, $W_{o}$(O{\sc
  i}1302.1)=1.69$\pm$0.07\AA\, $W_{o}$(C{\sc
  ii}$\lambda$1334.5)=0.32$\pm$0.07\AA\, $W_{o}$(Si{\sc
  iv}$\lambda$1393.4)=0.24$\pm$0.06\AA\ and $W_{o}$(Si{\sc
  iv}$\lambda$1402.8)=0.29$\pm$0.06\AA.  Since we can not constrain the
ionisation state or metallicity of the galaxy (we only have one
transition from each species), we adopt a simple approach and assume
that the ionisation and metallicity are comparable to that of cB58 at
$z=2.72$ which has an abundance of $\sim0.4$ solar \citep{Pettini02b}.
Assuming $log[X/H]$=-4.86 for Si{\sc ii}$\lambda$1260.4 and an
equivalent width of $W_{o}$(Si{\sc
  ii}$\lambda$1260.4)=0.30$\pm$0.07\AA\ (which corresponds to a column
density of $N_{SiII}\sim$2$\times$10$^{13}$\,cm$^{-2}$) we derive a
hydrogen column density of $\sim$2$\times$10$^{17}$\,cm$^{-2}$.
Although we caution this estimate comes with considerable uncertainty,
the column density estimated from the Ly$\alpha$ line profile, or
crudely using the UV ISM absorption lines appear, at least, to be
consistent within a factor $\sim$4$\times$.

Although there are uncertainties in the column density of the
outflowing shell, it is interesting to consider the energetics.  For a
normal stellar initial mass function, supernovae provide
$\sim$10$^{49}$\,erg per solar mass of stars (eg., \citealt{Benson03}),
thus for a galaxy with a star-formation rate of SFR=42$\pm$7$\Msolyr$,
$\sim$5$\times$10$^{50}$\,erg are available per year. We estimate the
distance of the shell from the galaxy by assuming that the linear size
of the region covered by the swept up shell is no larger than the
spatial extent as the galaxy (which seems reasonable given the velocity
structure of the outflow appears to follow the nebular emission line
velocity).  As noted, at a velocity of $\sim150\,\kms$ it takes 15\,Myr
to travel out to a distance of 2\,kpc; within this time, supernovae
will provide a total energy of E$_{SNe}\sim$5$\times$10$^{57}$\,erg.
Clearly there are a number of uncertainties in this estimate: not least
that the outflow may have decelerated from an initial higher velocity
which would reduce the time-scale.  Nevertheless, we can estimate the
mass of the outflow via: $M_{outflow} = AN_{HI}\times m_{H}/x$ where
$N_{HI}$ is the column density, $A$ is the area of the cone and $m_{H}$
is the atomic mass of Hydrogen.  Adopting the value of the observed
column density derived above
($N_{\rm{HI}}\sim\times10^{17-18}\,$cm$^{-2}$) and that the outflowing
material is uniform over an area of $\sim$10\,kpc$^{2}$, the total mass
of the outflow is $3\times10^{4}/x\Msol$, where $x$ is the neutral
H{\sc i} fraction.  The kinetic energy of the outflow is then
$E_K=\frac{1}{2}mv^{2}\sim10^{53}/x$\,erg.  Thus, the outflow is easily
energetically feasible even if the neutral fraction is as low as
0.01\%.  The low inferred mass loading and kinetic energy in the shell
is in stark contrast to the $z$=4.88 galaxy behind the lensing cluster
RCS0224-002 in which similar observations suggest a large scale,
bi-polar outflow surrounds the galaxy \citep{Swinbank07a}.  In this
galaxy, the outflow has a high mass loading (comparable or greater than
the star-formation rate) and appears to be located $>$30\,kpc from the
galaxy, escaping at a speed of up to 500\,km\,s$^{-1}$.  The strong
contrast between the two galaxies so far studied at these early times
therefore suggests strong diversity in the outflow energetics of young
galaxies at high-redshift, clearly illustrating the need for more
targets and follow-up to test the ubiquity and impact of outflows at
these early times.

\subsection{Comparison of Age and Mass Estimates}

It is useful to compare the various stellar and dynamical mass and age
estimates, since these reflect the build up of stellar mass in this
system.  First, the stellar mass estimated from the rest-frame
UV--near-infrared SED of the galaxy is much smaller than the dynamical
mass estimated from the dynamics.  Similar results have been found in
the $z\sim$2--3 population.  \citet{Erb06a} find $\sim$10\% of
star-forming systems with stellar masses which are approximately 10\%
of the dynamical mass.  These systems also tend to be the youngest;
$<$20\,Myr (e.g. HDF-BX1397 is an excellent example).  Thus, although
the stellar mass estimate appears small compared to the dynamical mass
(possibly suggesting a large gas reservoir makes up the rest of the
baryonic material), the galaxy does not appear to be at odds with
comparable observations of the galaxy population at $z\sim$2--3.

It is also interesting to note that the age estimated from the SED
($\sim$15\,Myr) is comparable to that derived from the instantaneous
star-formation rate (at a SFR of 42\,M$_{\odot}$\,yr$^{-1}$ it takes
just $\sim$17\,Myr to build a stellar mass of
7$\times$10$^{8}$\,M$_{\odot}$).  Finally, the dynamical age of the
outflow ($\sim$15\,Myr) is young, and consistent with a scenario in
which a starburst wind is in the process of accelerating material from
the galaxy into the halo and/or IGM.  Together, these mass and age
estimated are all suggestive of a young galaxy experiencing its first
major epoch of mass assembly.

\subsection{Properties of the Largest Star-Forming Regions}

\begin{figure}
\centerline{
  \psfig{figure=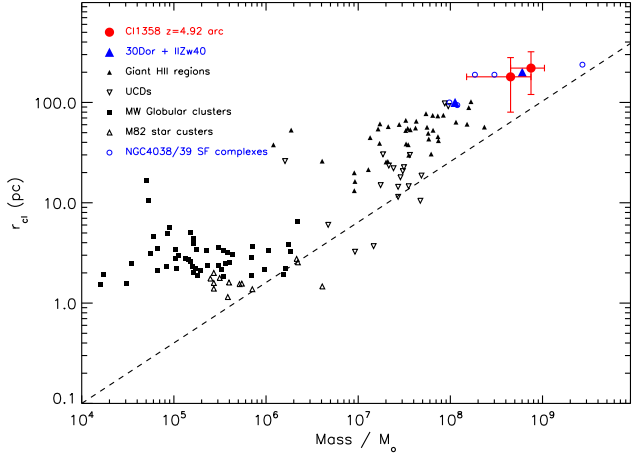,width=3.5in,angle=90}}
\caption{Mass versus size for H{\sc ii} regions in local galaxies
  compared to the two H{\sc ii} regions in MS\,1358+62 showing that the
  the two brightest star-forming regions are large, massive star
  clusters, but have comparable densities to H{\sc ii} regions found
  locally.  The squares denote Milky-Way globular clusters
  \citep{Harris96,Pryor93}, inverted open triangles represent
  Ultra-Compact Dwarfs \citep{Hasegan05,Evstigneeva07,Hilker07} and
  open triangles are M82 superstar clusters from \citet{McCrady07}. The
  solid triangles denote giant H{\sc ii} regions from
  \citet{Fuentes-Masip00}.  The dashed line shows the predicted cluster
  radius as a function of mass (after accounting for mass loss) from
  \citet{Murray09}.}
\label{fig:masssize}
\end{figure}

\begin{figure}
\centerline{
  \psfig{figure=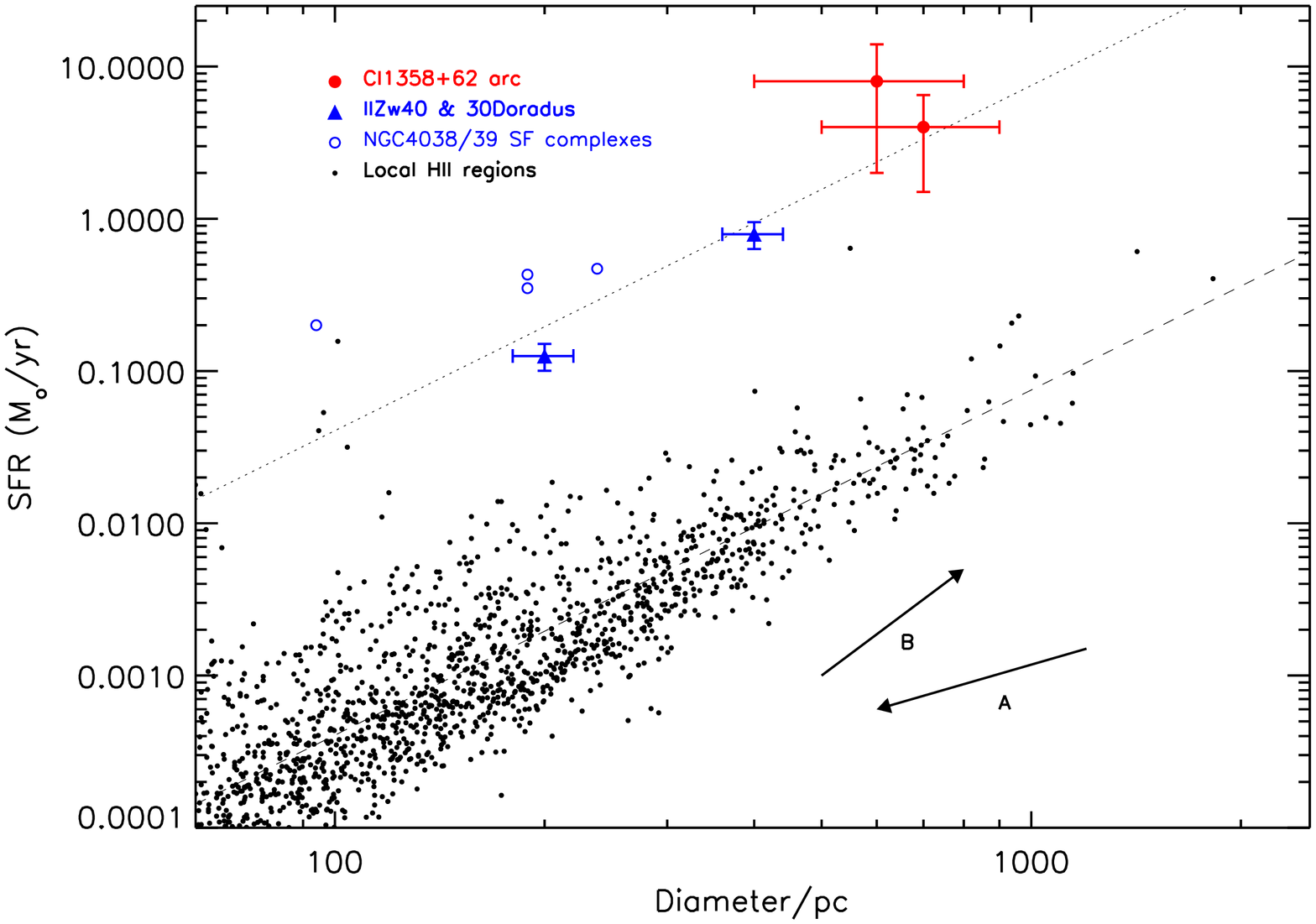,width=3.5in,angle=90}}
\caption{Correlation between size and star-formation rate from
  H$\alpha$ for H{\sc ii} regions in local galaxies.  The solid points
  denote the H{\sc ii} regions in local star-forming galaxies from
  \citet{Kennicutt88,Kennicutt03,Lee07}.  We also show the
  star-formation rate--size relation for 30 Doradus (the largest, local
  H{\sc ii} region in the LMC) as well as the giant starburst region in
  the nearby galaxy II\,Zw\,40 \citep{Vanzi08} and the Antennae
  \citep{Bastian06}.  The dashed line shows a fit to the H{\sc ii}
  regions in nearby spiral galaxies whilst the dotted line is the same
  but shifted in luminosity at a fixed size by a factor 100$\times$.
  The two H{\sc ii} regions in MS\,1358+62\,arc are shown by the solid
  red points where the ${\it x}$ error-bars denote the range of size
  estimates from the source-plane IFU and ACS reconstructions, and the
  y-error bar denotes the error in the flux measurement, lens modeling
  uncertainty and the range of possible [O{\sc ii}]/H$\alpha$ emission
  line flux ratios from \citet{Kewley04} for extreme ranges in
  metallicity and reddening.  The two H{\sc ii} regions in the $z=4.92$
  galaxy appear to be scaled up versions of the most luminous
  star-forming regions observed locally, even though they are observed
  when the Universe was less than 1\,Gyr old.}
\label{fig:sizeSFR}
\end{figure}

Owing to the magnification of the galaxy, the source-plane resolution
of the optical and near-infrared imaging we are also able to estimate
the masses, sizes and luminosities of the brightest star-forming (H{\sc
  ii}) regions within the galaxy to test whether the physical
conditions for star-formation at $z\sim5$ are very different to
galaxies today.  Using the results from \S\,3.1 we estimate the
brightest star-forming regions within the galaxy have diameters of
order $\sim$300--400\,pc from the {\it HST} and NIFS observations.
Using the velocity dispersions measured from the IFU observations this
suggests dynamical masses of $\sim$6--9$\times$10$^{8}$\,M$_{\odot}$
each and mass densities of $\sim$3000\,M$_{\odot}$\,pc$^{-2}$.  In
Fig.~\ref{fig:masssize} we show the correlation between size and mass
of H{\sc ii} regions (and starburst complexes) in local galaxies.
Clearly, the size density of the brightest star-forming regions in
MS\,1358+62\,arc are consistent with the most massive clusters seen in
star-forming galaxies.  Next we compare the luminosities and sizes of
the brightest H{\sc ii} regions to those locally.  In
Fig.~\ref{fig:sizeSFR} we show the correlation between size and
luminosity of local H{\sc ii} regions \citep{Kennicutt88,
  Gonzalez97,Kennicutt03,Lee07} and show the position of the two
brightest star-forming regions from MS\,1358+62\,arc.  As
Fig.~\ref{fig:sizeSFR} shows, the two brightest H{\sc ii} regions
appear to have star-formation activity which is substantially higher
(at a fixed size) than those typically derived from local star-forming
galaxies.

However, before we interpret this offset in detail, there are two
caveats which must be considered.  We first note that this comparison
relies on a conversion between [O{\sc ii}] (in the $z=4.92$ arc) and
H$\alpha$ locally.  However, as \citet{Kewley04} point out, the
conversion between the star-formation rate derived from H$\alpha$ and
[O{\sc ii}] never varies by more than a factor of two, even for extreme
ranges in metallicity and reddening.  Indeed, targeting the nebular
emission line ratios of $\sim$25 lensed galaxies, Richard et al.\ (2009
in prep) (see also \citealt{Pettini01}) derive a median
H$\alpha$/[O{\sc ii}] emission line flux ratio of H$\alpha$/[O{\sc
    ii}]=2.5$\pm$0.8 (median redshift of this sample $z=$2.4$\pm$0.4),
which is slightly larger, but consistent with the values locally
(H$\alpha$/[O{\sc ii}]=1.8$\pm$0.5; \citealt{Kewley04}).  Thus it seems
unlikely that there is a factor 100$\times$ increase in the [O{\sc
    ii}]/H$\alpha$ emission line ratio between $z=2$ and $z=5$.

Second, we caution that the isophotes used to extract H{\sc ii} regions
sizes and luminosities are very different at $z=$0 compared to those at
$z\sim5$ (even accounting for lensing).  Hence, to test whether we can
reliably compare H{\sc ii} region sizes and luminosities at
high-redshift to those locally, we make use of the extensive H$\alpha$
narrow-band imaging of local galaxies, in particular from SINGS
\citep{Kennicutt03} and 11\,HUGS \citep{Lee07}.  Using 175 H$\alpha$
narrow-band images from these samples we test the effect of redshifting
the $z=0$ galaxy images and re-extracting the size--luminosity
relation.  First, we use the the H$\alpha$ narrow-band imaging and
construct the size--luminosity relation for local galaxies using
isophotes ranging from 1$\times$10$^{-18}$ to
1$\times$10$^{-15}$\,erg\,cm$^{-2}$\,s$^{-1}$\,sq-arcsec.  We find
that, as the isophotal flux is increased, the lower-luminosity H{\sc
  ii} regions fall below the detection threshold, whilst only the cores
of the higher luminosity H{\sc ii} regions are identified.  This
results in the local relation remaining in tact, with fewer H{\sc ii}
regions identified.  In Fig.~\ref{fig:sizeSFR} we show a vector
(labeled 'A') which shows the typical direction a H{\sc ii} regions
follow as the extraction isophote is increased.

Next, we test the effect of artificially redshifting the imaging from a
spatial resolution of $\sim$10\,pc (0.5$''$ at 5\,Mpc) to a resolution
of $\sim$200\,pc and extract the H{\sc ii} regions at a fixed isophote.
This binning has the effect of merging H{\sc ii} regions, increasing
both the size and luminosity and in Fig.~\ref{fig:sizeSFR} we show
vector 'B' which denotes the track which H{\sc ii} regions follow as
the resolution is decreased.

The simulations show that essentially the size--luminosity relation
does not systematically shift as the images are redshifted, and we
therefore feel confident that any offset we measure in the
size--luminosity relation likely reflects a real difference their
properties (see also \citealt{Melnick00}.)  We note that it is entirely
possible that the star-forming regions in the $z=5$ galaxy represent an
amalgamation of multiple H{\sc ii} regions.  However, these simple
simulations suggest that they will remain over-luminous at a fixed
size.

Next, we examine the implication of identifying H{\sc ii} regions at
$z=4.92$ which are significantly more luminous than those found
locally.  At a fixed size of 400\,pc, local H{\sc ii} regions have
star-formation rates of order SFR=0.1\,M$_{\odot}$\,yr$^{-1}$, thus the
star-forming regions observed in the $z=$4.92 galaxy appear to be
substantially more luminous than those in local galaxies.  However, at
a fixed size, offsets of (up to) two orders of magnitude in
luminosities are observed in the most luminous nearby H{\sc ii} regions
and starburst galaxies
(eg. \citealt{Wilson95,English03,Vanzi08,Bastian06}).  For example,
30\,Doradus has a star-formation rate of
$\sim$0.15\,M$_{\odot}$\,yr$^{-1}$ and a size of $\sim$200\,pc whilst
the giant H{\sc ii} region in the starburst galaxy II\,Zw\,40 has a
star-formation rate of $\sim$1\,M$_{\odot}$\,yr$^{-1}$ within 400\,pc
(eg. \citealt{Wilson95,English03,Vanzi08}), although both of these
systems are low metallicity.  However, observations of massive
star-clusters in NGC\,4038/39 (the Antennae) also indicate young
($\sim$5\,Myr) clusters with star-formation rates much higher than
expected for typical star-forming galaxies \citep{Bastian06}.  Thus the
H{\sc ii} regions within this $z=4.92$ galaxy appear to be scaled up
versions of the most extreme H{\sc ii} regions observed in the local
Universe.  It is unclear whether this is due to the starburst mode, or
low metallicity, although it is interesting to note that the most
recent models of galaxy formation suggest that star-formation at high
redshift may be driven by strongly unstable and fragmenting disks
(eg. \citealt{Dekel09}) resulting in localised starbursts.  Thus, the
intense star-formation observed in this galaxy will significantly
effect the surrounding ISM both in distributing metals and in adding
turbulence.

\section{Discussion \& Conclusions}

After correcting for lensing magnification factor 12.5$\pm$2.0$\times$,
the rest-frame UV/optical morphology appears extended over
$\sim$4\,kpc, but is dominated by approximately five star-forming
regions.  In the source plane the galaxy has an $I$-band magnitude
consistent with L$^{*}$ LBGs at z$\sim$5, and an intrinsic
star-formation rate of SFR$_{[OII]}=$42$\pm$8\,M$_{\odot}$\,yr$^{-1}$.
Using rest-frame UV-to-$H$-band SED we derive stellar mass of
M$_{\star}=$(7$\pm$2)$\times$10$^{8}$M$_{\odot}$, which is consistent
with estimates of the stellar masses of LBGs at this epoch
\citep{Verma07,Stark09,McLure09}.  The SED is also consistent with
being dominated by a young stellar population of age of 14$\pm$7\,Myr.
Indeed, with a star-formation rate of $\sim$42\,M$_{\odot}$\,yr$^{-1}$,
it takes just $\sim$17\,Myr to build this stellar mass.  Together, this
suggests this galaxy is in the first major epoch of mass assembly.

The source-plane morphology is clumpy with at least five star-forming
regions, separated by up to 4\,kpc in projection.  The brightest
regions are not centrally concentrated, and the galaxy shows no signs
of a central component.  Using the spatially resolved spectroscopy
around the [O{\sc ii}] emission, we derive a velocity gradient of
$\Delta v$=180$\pm$20\,km\,s$^{-1}$ across 4\,kpc in projection,
suggesting a dynamical mass of
M$_{dyn}$=3$\pm$1$\times$10$^{9}$M$_{\odot}${\it
  csc$^2$(i)}\,M$_{\odot}$.  We also derive a specific star-formation
rate (SFR/M$_{\star}$) of SSFR=2.4$\pm$0.5$\times$10$^{-8}$\,yr$^{-1}$,
consistent with the most actively star-forming, low mass galaxies at
these early times \citep{Fueler05}.

Using the source-plane images and [O{\sc ii}] emission line maps, we
measure sizes and masses for the two brightest star-forming regions,
deriving masses of M$_{cl}$=6--9$\times$10$^{8}$\,M$_{\odot}$ and
individual star-formation rates of
SFR$_{[OII]}\sim$10\,M$_{\odot}$\,yr$^{-1}$.  A comparison to local
H{\sc ii} regions/populations shows that these are comparably massive
and dense as those found in local star-forming galaxies.  However, the
implied star-formation rate individually for these H{\sc ii} regions is
$\gsim$100$\times$ brighter that found at a fixed size.  Such high
star-formation rates (at a fixed size) have been observed locally:
e.g. 30 Doradus and II\,Zw\,40 both have star-formation rates which are
$>$100$\times$ larger than expected given their sizes.  Thus, the H{\sc
  ii} regions within this $z=4.92$ galaxy appear to be scaled up
versions of the most extreme regions observed in the local Universe,
yet observed when the Universe was $\lsim$1\,Gyr old.  Could the
increased luminosity within H{\sc ii} regions reflect real difference
in the mode of star-formation within massive star-forming complexes?
\citet{Murray09} suggest that young, massive ($>$10$^{7}$\,M$_{\odot}$)
star-forming clusters are optically thick to far-infrared radiation
resulting in high gas temperatures, and hence in higher Jeans masses.
Indeed, for a star cluster with mass $>$10$^{8}$\,M$_{\odot}$ (as
observed in this $z=4.92$ galaxy) the predicted Jeans mass is
$\sim$12\,M$_{\odot}$.  This 'top-heavy' IMF has the effect of
increasing the fraction of OB stars per H{\sc ii} region and hence
increases the light-to-mass ratio (see the discussion in
\citealt{Murray09}).  However, since no compelling evidence for
evolution in the IMF has yet been found at either low (or high)
redshift, we view this scenario as suggestive, at best.  Nevertheless,
clearly spatially resolved spectroscopy of the nebular emission lines
on scales comparable to H{\sc ii} regions would provide crucial
diagnostics of the physics of star-formation in the young Universe.
Indeed, this intense star-formation will significantly effect the
surrounding ISM both in distributing metals and in adding turbulence.
As such, the mechanical energy input might explain the higher turbulent
speeds (hence large v/$\sigma$ values) observed in primitive disks
\citep[e.g.][]{Stark08,Law09,ForsterSchreiber09,Genzel09,Lehnert09}.

Finally, we compare the nebular emission line dynamics with spatially
resolved Ly$\alpha$ and rest-frame UV-ISM lines and find that this
galaxy is surrounded by a galactic scale outflow.  The velocity
gradient observed in the nebular emission is mirrored (but
systematically offset) in the Ly$\alpha$ emission and UV-ISM absorption
across $\sim$4\,kpc, suggesting that the outflow is young ($<$15\,Myr)
and has yet to decouple from the star-forming regions and escape from
the galaxy disk.  Although crude, the estimated mass loading in
the outflow is very small, and the kinetic energy provided by SNe from
the observed star-formation rate is easily enough to drive the wind.

Overall, these observations provide unique insights into the
distribution of star-formation, dynamics, interaction between
star-formation and outflow energetics and even properties of the
largest H{\sc ii} regions within a young galaxy seen less than 1\,Gyr
after the big-bang on scales of just $\sim$200\,pc.  The combination of
the large number of high-redshift gravitationally lensed galaxies
identified by {\it HST} and adaptive optics assisted integral field
spectrographs on eight and ten meter telescopes should finally begin to
make such critical observational studies common-place, testing the
route by why early systems assemble their stellar mass, their mode of
star-formation and how they ultimately develop into galaxies like the
Milky-Way.

\section*{acknowledgments}

We gratefully acknowledge an anonymous referee for a very constructive
report which significantly improved the content and clarity of this
paper.  We also gratefully acknowledge the Gemini staff for granting SV
time for these observations.  We would like to thank Janice Lee for
very useful discussions and for allowing us to use the 11\,HUGS
narrow-band imaging prior to publication and Marijn Franx for useful
discussions and providing us with the rest-frame UV spectroscopy.  AMS
gratefully acknowledges a Sir Norman Lockyer Royal Astronomical Society
Fellowship and a Royal Society travel grant and DPS acknowledges an
STFC fellowship.  The Gemini observations were carried out as part of
program GN-2006-SV-130, based on observations obtained at the Gemini
Observatory, which is operated by the Association of Universities for
Research in Astronomy, Inc., under a cooperative agreement with the NSF
on behalf of the Gemini partnership: the National Science Foundation
(United States), the Particle Physics and Astronomy Research Council
(United Kingdom), the National Research Council (Canada), CONICYT
(Chile), the Australian Research Council (Australia), CNPq (Brazil) and
CONICET (Argentina).

\bibliographystyle{apj}
\bibliography{/Users/ams/Projects/ref}
\bsp

\end{document}